# Multivariate prediction of mixed, multilevel, sequential outcomes arising from in vitro fertilisation

Jack Wilkinson*, Andy Vail, Stephen A Roberts

*Centre for Biostatistics, Institute of Population Health, Manchester Academic Health Science Centre, University of Manchester, Manchester, UK*

*corresponding author: jack.wilkinson@manchester.ac.uk

**Abstract.** In vitro fertilization (IVF) comprises a sequence of interventions concerned with the creation and culture of embryos which are then transferred to the patient's uterus. While the clinically important endpoint is birth, the responses to each stage of treatment contain additional information about the reasons for success or failure. As such, the ability to predict not only the overall outcome of the cycle, but also the stage-specific responses, can be useful. This could be done by developing separate models for each response variable, but recent work has suggested that it may be advantageous to use a multivariate approach to model all outcomes simultaneously. Here, joint analysis of the sequential responses is complicated by mixed outcome types defined at two levels (patient and embryo). A further consideration is whether and how to incorporate information about the response at each stage in models for subsequent stages. We develop a case study using routinely collected data from a large reproductive medicine unit in order to investigate the feasibility and potential utility of multivariate prediction in IVF. We consider two possible scenarios. In the first, stage-specific responses are to be predicted prior to treatment commencement. In the second, responses are predicted dynamically, using the outcomes of previous stages as predictors. In both scenarios, we fail to observe benefits of joint modelling approaches compared to fitting separate regression models for each response variable.

1. **Background and motivation**

In vitro fertilization (IVF) is a complex multistage procedure for the treatment of subfertility. Typically, a 'cycle' of IVF begins with the administration of drugs to stimulate the patient's ovaries and promote the release of oocytes (eggs). The oocytes are collected from the patient and are then fertilised either by mixing or injecting them with sperm. The resulting embryos are cultured for several days. Finally, one or more of the best embryos are selected for transfer to the woman's uterus, where it is hoped that they will implant and develop into a healthy baby. Treatment may fail at any stage of the cycle (if no oocytes are recovered from the ovaries, no good quality embryos are produced, or those transferred do not implant), in which case the subsequent stages are not undertaken.

The sequential nature of IVF means that the patient's response can be measured at each stage of the treatment (1): the stimulation of the ovaries can be evaluated by the number of oocytes collected; the fertilization and culture stages can be evaluated by the number and quality of embryos produced; and the success of the transfer procedure can be evaluated according to whether or not a child is born as a result. Figure 1 displays a schematic of the IVF cycle. A recent review of outcome measures used in IVF RCTs showed that there is considerable interest in these 'intermediate' or 'procedural' outcomes of IVF; 361 distinct numerators were identified, and the median (IQR) number of distinct outcomes reported per trial was 11 (7 to 16) (2).

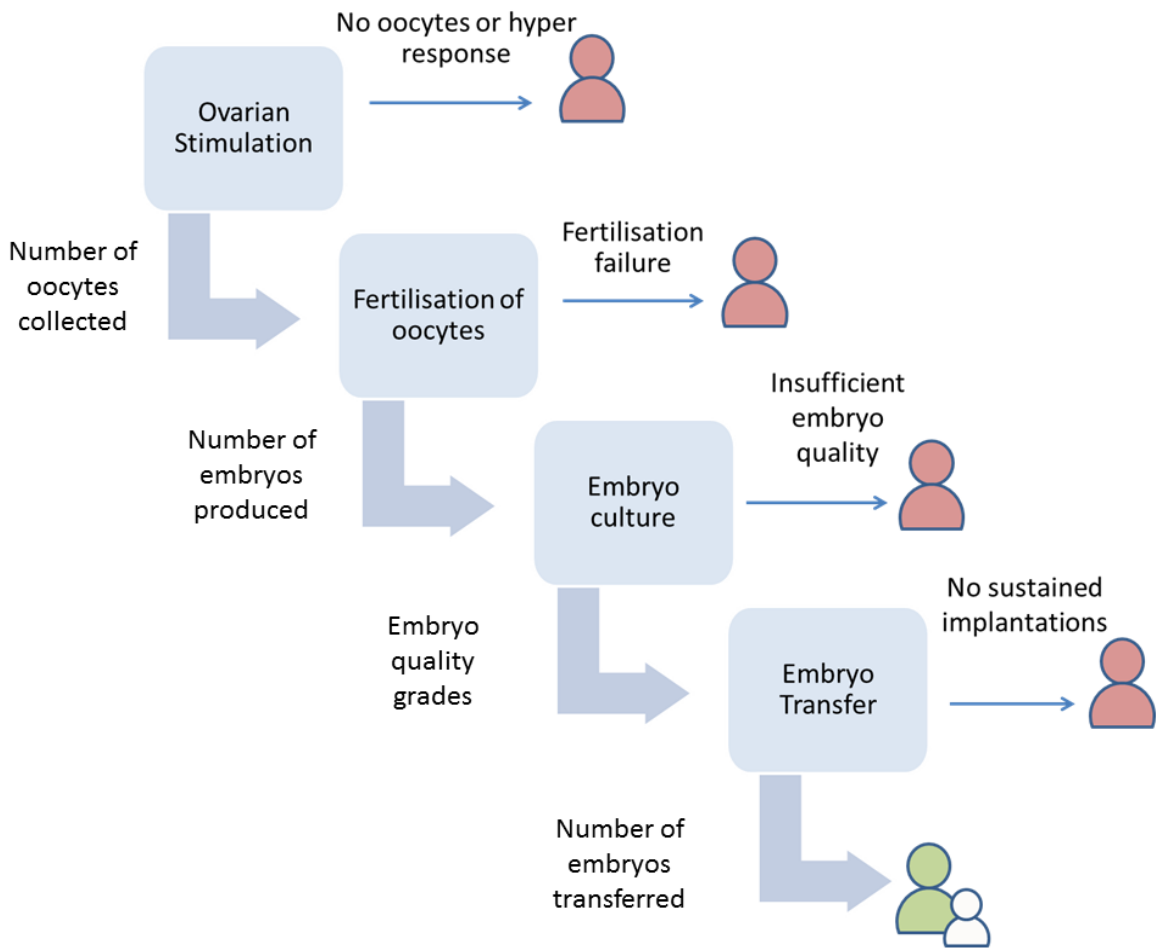

Figure 1: Schematic of the IVF cycle

The interest in procedural outcomes in IVF research is not surprising. While the most relevant measure of success for patients is the birth of a child (3), (1), (4), the procedural outcomes have material implications for safety and for the scope to undertake additional treatment attempts if the first turns out to be unsuccessful. For example, it is important to avoid excessive response to ovarian stimulation, since this may result in ovarian hyperstimulation syndrome, resulting in hospitalisation or, in rare instances, death (5). And, the availability of spare good quality embryos, which may be cryopreserved, is a requirement to undergo additional embryo transfers without a further round of ovarian stimulation. Recognising the potential value of information contained in procedural outcomes, two approaches for the analysis of multistage IVF data have recently been proposed (6, 7). The first is a discrete time-to-event approach that treats the stages of the IVF cycle as a series of 'failure opportunities' (6). Each woman's response data then comprise a vector of binary indicator variables denoting whether treatment failed at this stage, or proceeded to the next. The second treats the stage of the cycle reached by the patient as an ordinal response, and models this using continuation ratio regression (7). Both of these approaches allow us to model the relationship between baseline treatment and patient characteristics and IVF response, while preserving the sequential nature of the data. Both share similar limitations, however. In particular, both treat the responses at each stage as dichotomous 'success or failure' events. This wastes a great deal of information, since it is more informative to measure the number of oocytes obtained from the ovaries than merely whether a sufficient quantity were available to enable the cycle to continue; and it is more informative to measure the quality of any embryos obtained than merely whether there were any available for transfer. These methods are also incapable of accommodating outcomes defined at different levels of a multilevel structure; some outcomes (eg: number of oocytes) may be defined for each patient, while others (eg: embryo quality) are defined for the patient's individual embryos. In addition, while these methods allow for differential effects of covariates at each stage through the inclusion of interaction terms, they do not allow for different covariates to be included as predictors for the different stage-specific responses.

While methods for the analysis of sequential IVF data exist therefore, it remains to identify techniques capable of incorporating the variety of outcome types encountered in this setting, and moreover responses which are defined at different levels of a two-level data structure (embryos and patients). This includes counts of oocytes, ordinal embryo quality scales, binary birth indicator variables, and so on. Methods for the analysis of multivariate responses of mixed outcome types are hardly new (eg: (8)) but have received considerable attention in recent years (see de De Leon and Chough (9) for a comprehensive collection of the state of the art). While much of this work has focussed on the joint analysis of time-to-event and longitudinal response data (see reviews by Gould, Boye (10) and Tsiatis and Davidian (11)), approaches capable of accommodating different combinations of outcome types have been described (12); (13), (14), (15), (16), (17), (18). Typically, these involve the inclusion of shared (12), (13), (14), (13), (16) or otherwise correlated (15), (18) latent variables in 'submodels' for the different response variables. These latent variables accommodate dependency between the response variables in the model. A further attractive feature of latent variable approaches is that they can be used to jointly model responses measured at different levels of a multilevel data structure (18), (17). These methods do not appear to have been discussed in the context of multistage treatments however.

Besides a multivariate approach, an alternative strategy for the prediction of IVF outcomes would be to explicitly model each of the patient's stage-specific responses using a series of unrelated regression equations. Given the relative simplicity of this approach, the pertinent question is whether there would be any material advantage to adopting the more complex, joint modelling framework. Some recent work has suggested advantages of multivariate approaches over fitting of separate models (19), albeit in a simpler context (two binary outcomes, measured at the same level) compared to the scenario considered here. A further consideration relates to the possibility of including stage-specific responses as predictors of downstream outcomes, since the outcome at each stage is likely to strongly predict what will happen next. The decision to include or exclude stage-specific responses as predictors of downstream events changes the function of the prediction model, since a model taking future events as input cannot be used prior to treatment. Incorporating the stage-specific outcomes would move the function of the model towards dynamic prediction, conditional on information accrued up until that point in the treatment cycle. This sort of approach could prove useful when making stage-specific decisions based on what has happened up to that point.

Stage-specific outcomes could be included as predictors in a series of distinct, sequential regression models for the response variables, or within the submodels comprising a multivariate or joint model. The latter approach would then resemble the endogenous treatment models employed in the econometrics literature (20), or multiprocess models that have been employed in education research (21), although these applications have focussed on causal inference, rather than prediction.

In this paper, we develop and illustrate methodology for the prediction of multistage IVF data, with mixed response types (count, ordinal, and binary) defined at different levels of a two-level data structure (patients and embryos), using an application to routinely collected data from a large reproductive medicine unit. We consider both multivariate models and approaches based on fitting distinct regression models for each outcome variable. We also consider models excluding and including stage-specific responses as predictors of downstream events, which would respectively represent models to be deployed prior to treatment, or dynamically.

In section 2, we describe the joint modelling approach. In section 3 we illustrate the use of the methods with an application to a routine clinical database. This is followed by a discussion in section 4.

## 2. Models

Here we describe a joint modelling approach for the analysis of multistage IVF data. The approach includes distinct submodels for each of the response variables considered in the cycle. We include six response variables for patient $j = 1,…,n$ and their embryos $i = 1,…,n_j$, and hence six submodels, in the current presentation: the number of oocytes (eggs) obtained from ovarian stimulation (a count, $y_j^O$); the fertilisation rate when the oocytes are mixed with sperm ($y_j^M$); two measures of embryo quality (cell evenness and degree of fragmentation $y_{ij}^E$ and $y_{ij}^F$, both measured using ordinal grading scales); an indicator denoting whether one or two embryos were transferred to the patient (denoted by a binary variable $y_j^D$) and another ($y_j^L$) indicating whether or not the transfer of embryos resulted in the live birth of one or more babies (a live birth event, or LBE) (Fig.1). These are listed in temporal order, with the exception of the two embryo quality scales, which are coincident. We include the decision to transfer two embryos (known as double embryo transfer, or DET) in the model because it is an important predictor of transfer success which is partially determined by the outcomes of the earlier stages. A second feature is that once a patient has dropped out of the cycle, she does not appear in the submodels corresponding to the downstream responses. In the following, we do not deal with the possibility that each patient may undergo multiple cycles of IVF, noting that the models could be extended to three levels (embryos nested within cycles nested within women) by adding additional random scalar terms (22).

*2.1 Joint model*

The joint model requires the use of latent variable representations for the various submodels constituting the larger model. Each patient *j* has associated vectors of responses $\mathbf{y}_j = (y_j^O, y_j^M, y_{ij}^E, y_{ij}^F, y_j^D, y_j^L)$ and of underlying latent variables $\mathbf{z}_j = (z_j^O, z_j^M, z_j^E, z_j^F, z_j^D, z_j^L)$. Both of these vectors may be partially observed due to drop-out or outright failure before completion of the treatment. We then posit a multivariate Normal distribution for the latent variables, and estimate the elements of the correlation and variance-covariance matrices. We prefer to use distinct latent variables in each submodel to an approach based on a common latent variable which is scaled by factor loadings in each submodel (eg: (13), (12)), as the linearity assumption required for the latter is too restrictive for present purposes (14). The submodels for each stage are presented below, followed by the multivariate distribution of latent variables.

*2.1.1    Stimulation phase submodel*

For patient *j*, we assume the number of oocytes (eggs) obtained $y_j^O$ follows a Poisson distribution and model the log of the rate parameter $\lambda_j^O$ in the usual way:

$$\log(\lambda_j^o) = \mathbf{X}_j^o \boldsymbol{\beta}^o + z_j^o \qquad (1)$$

where $\mathbf{X}_j^o$ is a row-vector of cycle-level covariates for patient *j*, $\boldsymbol{\beta}^o$ is a corresponding vector of regression parameters and $z_j^o$ is a patient-specific latent variable that captures overdispersion in the oocyte yield. This submodel is fitted to all patients who start the cycle.

*2.1.2    Fertilisation submodel*

We model the number of embryos obtained when oocytes are mixed with sperm $y_j^M$ in terms of its rate parameter $\lambda_j^M$, again using a Poisson submodel:

$$\log(\lambda_j^M) = \log(y_j^O) + \boldsymbol{X}_j^M \boldsymbol{\beta}^M + z_j^M \tag{2}$$

where $\boldsymbol{X}_j^M, \boldsymbol{\beta}^M$ and $z_j^M$ are analogous to the corresponding terms in the stimulation model. We now include an offset term corresponding to the logarithm of the number of oocytes obtained in the linear predictor. This submodel is fitted to all patients who have oocytes mixed with sperm. In some cycles, the number of oocytes mixed with sperm is less than the number obtained, so there is an implicit assumption in the model that any oocytes which were not mixed could not have been successfully fertilized. The assumption is reasonable, since the decision not to mix an oocyte with sperm is almost always based on the fact that the oocyte has been identified as being degenerate.

*2.1.3 Embryo quality submodels*

We include two measures of embryo quality; cell evenness ($y^E$) and degree of fragmentation ($y^F$). These are ordinal 1 to 4 grading scales measured at the level of individual embryos. We model these using cumulative logit submodels. For embryo *i* (where *i* = 1,2,…,$n_j$) nested in patient *j* we have, for *k* = 1,2,3:

$$\begin{aligned}\text{logit}(\gamma_{kij}^E) &= \alpha_k^E - \boldsymbol{X}_{ij}^E \boldsymbol{\beta}_k^E - z_j^E \\ \text{logit}(\gamma_{kij}^F) &= \alpha_k^F - \boldsymbol{X}_{ij}^F \boldsymbol{\beta}_k^F - z_j^F\end{aligned} \tag{3}$$

where $\boldsymbol{X}_{ij}^{E}$ and $\boldsymbol{X}_{ij}^{F}$ are row-vectors of covariates, $\boldsymbol{\beta}_{k}^{E}$ and $\boldsymbol{\beta}_{k}^{F}$ are vectors of regression coefficients which may vary across the levels of *k* (relaxing the proportional odds assumption), and $z_j^E$ and $z_j^F$ are patient-level random effects (latent variables) which are identified due to the clustering of embryos within patients. $\gamma_{kij}^{E}$ and $\gamma_{kij}^{F}$ are cumulative probabilities of embryo *i* in patient *j* having a grade of *k* or lower for evenness and fragmentation degree respectively and $\alpha_k^E$ and $\alpha_k^F$ are threshold parameters, corresponding to the log-odds of the embryo having grade *k* or lower. The negative signs in these equations are used so that positive effects of a covariate may be interpreted as increasing the ordinal measure. These submodels are fitted to all embryos.

*2.1.4 Double embryo transfer submodel*

In order to jointly model the binary response DET, denoting the number of embryos transferred, with the other response variables, we use a latent variable representation of a probit regression model (23). Let $y_j^D = 1$ or 0 if patient *j* does or does not have DET, respectively. We define $y_j^{D*}$ as a latent continuous variable underlying the binary $y_j^D$, such that:

$$y_j^D = \begin{cases} 1 \ if\ y_j^{D*} \geq 0 \\ 0 \ if\ y_j^{D*} < 0 \end{cases} \quad (4)$$

A linear regression submodel for the latent $y_j^{D*}$ is then used to estimate covariate effects:

$$y_j^{D*} = \boldsymbol{X}_j^D \boldsymbol{\beta}^D + z_j^D \quad (5)$$

$$z_j^D \sim N(0,1)$$

where $X_j^D$ is a row-vector of patient-level covariates and $\beta^D$ is a vector of regression coefficients. Fixing the variance of $z_j^D$ to be 1 is mathematically equivalent to specifying a probit model for the probability that a patient will have DET. We use $z_j^D$ to link the DET submodel to the others.

### 2.1.5 Live birth event submodel

As for DET, we use a latent probit representation for $y_j^L = 1$ or 0 corresponding to whether or not LBE obtains, with an underlying latent variable $y_j^{L*}$:

$$y_j^L = \begin{cases} 1 \text{ if } y_j^{L*} \geq 0 \\ 0 \text{ if } y_j^{L*} < 0 \end{cases} \tag{6}$$

Again, a linear regression submodel for the latent $y_j^{L*}$ is then used to estimate covariate effects:

$$y_j^{L*} = X_j^L \beta^L + z_j^L \tag{7}$$

$$z_j^L \sim N(0, 1)$$

with row vector of patient-level covariates $X_j^L$ and vector of regression coefficients $\beta^L$. The error term $z_j^L$ again has a variance of 1, and is used to link the LBE submodel to the others. The DET and LBE submodels are fitted to patients who undergo the transfer procedure.

### 2.1.6 Covariates

Different covariates may be included in each of the covariate vectors $X_j^O, X_j^M, X_{ij}^E, X_{ij}^F, X_j^D, X_j^L$. If interest is in dynamically predicting the outcome of subsequent stages, conditional on responses at previous stages, this could include any of the response variables $y_j$ occurring prior to this stage. For example, the number of fertilised eggs could be included in the submodels for embryo evenness and fragmentation, or these measures of embryo quality could be included in submodels for DET and LBE.

When including outcomes as covariates in a joint model, model identification can pose a challenge (24), (25). Standard strategies include fixing parameters in the model (for example, fixing elements of the latent correlation matrix to be zero) and including non-identical sets of covariates in the submodels. When these approaches are used in a causal inference setting, this requirement translates to including instrumental variables in the submodels (26), (21), (27). Where our goal is prediction rather than causal inference however, these strong structural assumptions are not required.

### 2.1.7 *Latent variable distribution*

We specify a multivariate Normal distribution for the latent variables to connect the submodels, with variances of 1 for the probit DET and LBE submodels:

$$\begin{bmatrix} z_j^O \\ z_j^M \\ z_j^E \\ z_j^F \\ z_j^D \\ z_j^L \end{bmatrix} \sim MVN \left( \begin{bmatrix} 0 \\ 0 \\ 0 \\ 0 \\ 0 \\ 0 \end{bmatrix}, \begin{bmatrix} \theta_O^2 & \eta_1 \theta_O \theta_M & \eta_2 \theta_O \theta_E & \eta_3 \theta_O \theta_F & \eta_4 \theta_O & \eta_5 \theta_O \\ . & \theta_M^2 & \eta_6 \theta_M \theta_E & \eta_7 \theta_M \theta_F & \eta_8 \theta_M & \eta_9 \theta_M \\ . & . & \theta_E^2 & \eta_{10} \theta_E \theta_F & \eta_{11} \theta_E & \eta_{12} \theta_E \\ . & . & . & \theta_F^2 & \eta_{13} \theta_F & \eta_{14} \theta_F \\ . & . & . & . & 1 & \eta_{15} \\ \eta_5 \theta_O & \dots & \dots & \dots & \dots & 1 \end{bmatrix} \right) \quad (6)$$

We note that this framework estimates the relationships between patient and embryo-level responses.

## 3  Application of the methods to routinely collected IVF data.

### 3.1  Fit to St Mary's Data

We illustrate the methods in an application to a routine clinical dataset obtained from St Mary's Hospital Department of Reproductive Medicine, Manchester, England. Our primary aims were to establish the feasibility of the multivariate approach, to investigate whether the joint approach offered any apparent benefit compared to fitting separate regression models to each response variable, and to elucidate considerations specific to sequential outcome prediction. The dataset includes 2962 initiated IVF treatments undertaken by 2453 women between 2013 and 2015, including quality data on 12,911 embryos. For present purposes, we ignore the fact that some women underwent multiple cycles, noting that the current models could be extended to a three-level setting to account for this additional source of clustering (22). Characteristics of the treatment cycles in the dataset are presented in Table 1. We considered two settings. In the pre-treatment setting, we included only covariates which would be available prior to treatment commencement (Table 2). In the dynamic setting, we included stage-specific responses as covariates in submodels for downstream events (Table 2). We included age and partner age in all of the submodels as linear covariates (although nonlinearities should be considered in any genuine application). Both variables were standardised (subtracting mean and dividing by a standard deviation) for fitting; for reporting, the corresponding model coefficients were converted to the original scale. The models are flexible enough to allow different covariates to be included in different submodels; we include attempt number in the number of oocytes and DET submodels, pooling 4th and 5th attempts due to small numbers in these categories. In the embryo evenness and fragmentation submodels, we also include an indicator variable denoting whether the egg was fertilized by injecting it with sperm, or by mixing *in vitro*; the method of fertilisation to be used is typically determined prospectively. We suppose that covariate effects are constant across the levels of the ordinal embryo responses (proportional odds), although the methods can accommodate non-proportionality. In the dynamic setting, we additionally included the actual (as opposed to predicted) number of eggs and fertilisation rate as covariates in all downstream submodels, and include embryo evenness and fragmentation, averaged over all of the patient's embryos, as covariates in the DET and LBE submodels.

In both settings, we fitted a joint model, and separate regression models to each response variable.

### 3.2 Predicting sequential outcomes and assessing performance

The fitted models can be used to make sequential predictions about the IVF cycle, which can be plotted against data to visually evaluate calibration. Here, we plot the posterior predictive distribution against the training data, noting that this does not constitute validation of the models. Were a new dataset to be used instead, this would represent an assessment of mean calibration (28) for the multivariate responses. We also note that interest here is not in model fit, which could be assessed by calculating predictions including the patient-level random effects from the fitted models. Because we only have recourse to measured variables when making predictions for new patients, and not to unmeasured sources of heterogeneity, we draw predictions without incorporating the patient-specific random terms in each submodel. We note that when making predictions for new patients, it may be necessary to integrate over the random effects in order to achieve good calibration (29).

When doing this, it is important to pay attention to the denominator we use at each stage, since the multistage nature of IVF presents several options in this regard, and this is a recurring source of error in analysis and interpretation of data in the field (30) (31) (2) (32). For the patient-level outcomes (number of oocytes, number of embryos, double embryo transfer, live birth) the first option is to perform calculations including only the subset of participants reaching each stage (for example, predicting the number of embryos only in people who had any oocytes, or predicting whether live birth obtains only in the subset of patients who reached the stage of embryo transfer). The second option is to perform calculations including all patients who started the cycle. In this case, we would define patients who had no oocytes to have no embryos, and similarly define anyone for whom treatment failed prior to embryo transfer not to have had a live birth. The second option corresponds to the choice to adopt a treatment policy estimand, to use the terminology of recent work on targets of inference (33, 34). However, this second option does not apply to the embryo-level quality outcomes (evenness and fragmentation), since these measures are not defined if no embryos exist. Here, we use the 'per cycle started' denominator for the patient-level outcomes, but calculate embryo gradings on the condition that the number of embryos is not zero.

Further differences arise between the pre treatment and dynamic setting. In the former scenario, predictions of stage-specific outcomes are made before the start of the cycle, using only information available at that time. Using the 'per cycle started' denominator, this means that we predict at which stage treatment will fail (if it does), and assign failure outcomes at the subsequent stages. For example, if we predict that the patient will have no oocytes, we deterministically assign that patient to have no embryos (and hence no embryo quality gradings), no embryo transfer, and no live birth. In the dynamic setting, we use the information accrued up to that stage to make predictions, including the observed outcomes at previous treatment stages. This means that the observed (rather than predicted) number of oocytes can be used when predicting the number of embryos, and that both of these can be used when predicting quality of those embryos (which only becomes apparent several days later), and so on.

One motivation for multivariate prediction models is to predict the occurrence of particular combinations of outcome. For example, it is possible to use the pretreatment models to make predictions about whether cycles will have safe ovarian stimulation outcomes (for example, yielding fewer than 15 oocytes, as higher may prove hazardous to the patient) while also resulting in a live birth. Here, we use the posterior predictive distribution for this purpose. If interest is in predicting the prevalence of safe, successful outcomes in a population, rather than for individual patients, it is more appropriate to include the random effects from the various submodels when making the prediction.

### 3.3 Fitting the models

We use the R (35) implementation of the Bayesian software Stan (36) to fit the models. While the benefits (or drawbacks, depending on one's perspective) of Bayesian methods have been well rehearsed, our use of this software is primarily driven by pragmatism; the software is flexible and can accommodate complex multilevel models without the need to author custom sampling algorithms. We place weak Normal $(0,1000^2)$ priors on the regression parameters in the submodels, with the exception of those included in the latent probit submodels (that is, those corresponding to DET, LBE). Given the fact that the latent responses in these submodels have a variance of 1, we place Normal $(0, 2^2)$ priors on the regression parameters. These can be considered to be weakly informative prior distributions which improve efficiency in fitting the model by restricting the sampler to realistic values for these parameters (37). We place weakly informative Cauchy (0,2.5) priors on the free variance parameters. Finally, we use an LKJ prior distribution for the latent correlation matrix, which is uniform over all possible correlation matrices (38). We run the samplers for between two and four thousand iterations in each case, using three chains. We check convergence using the Gelman-Rubin convergence diagnostic (39) and using traceplots. In practice, we note that fitting the joint model with outcomes included as covariates took several days on an Intel Core i7-4810MQ 2.8 GHz processor with 16 GB of RAM. Stan and R code is available at https://osf.io/pmrn3/ .

### 3.4 Results and interpretation

3.4.1 *Model coefficients*

Table 2 shows the estimates from the four fitted models. Coefficients differed between the two pre treatment models, although generally differences were not dramatic. The most notable differences between the pre treatment joint and separate fits were in the intercepts for the embryo evenness and fragmentation submodels, and the coefficient for sperm being injected into the egg in the fragmentation submodel, which was closer to the estimates from the dynamic models in the joint approach. 95% CIs were not consistently narrower in the joint model compared to the separate model fits, and were never substantially so. Although perhaps of lesser importance, the statistical significance of the coefficients did not change between the separate and joint pre treatment models.

For the dynamic models, differences in estimates were often more dramatic, particularly for those corresponding to upstream outcome variables included as covariates, in terms of magnitude, direction, statistical significance, and precision. Notably, the joint model estimates for these covariates were considerably less precise compared to those obtained from separate fits. By analogy to selection models used in causal inference, which possess some similarities to the joint dynamic model, we speculate that strong, distinct predictors of outcome would need to be included in each submodel to overcome imprecision here; the analogous requirement in the causal inference context is for strong instrumental variables. Larger sample size may also be necessary.

*3.4.2 Assessing prediction of sequential outcomes*

Sequential predictions obtained from the models are presented in Figure 2. Note that, given the small number of covariates included in this example, the pre treatment models did not 'predict' any of the outcomes well without including the random effects, even in the training data (columns a and b). The joint pre treatment model did not represent a clear improvement over the separate fits. In the dynamic setting (columns c, d), including the outcomes of earlier treatment stages did appear to result in improved mean calibration of the number of embryos, DET and live birth, although again no benefit of a joint approach was apparent, and the joint model appeared to be worse for DET. However, RMSE did not vary across the models for either oocytes (6) or embryos (2.9). AUC (95% CIs) for both pre treatment models were identical, 0.58 (0.55 to 0.60) for DET and 0.55 (0.53 to 0.58) for LBE. For the dynamic separate and joint models AUCs (95% CI) were 0.64 (0.62 to 0.66) and 0.58 (0.55 to 0.60) for DET and 0.60 (0.58 to 0.63) and 0.47 (0.44 to 0.49) for LBE, with the joint model frequently predicting slightly higher probabilities in those who didn't have live births compared to those who did.

Using the median, 2.5 and 97.5 percentiles from the posterior predictive distribution obtained from the pre treatment models, 28% (26 to 31) were predicted to have a safe but successful outcome (fewer than 15 oocytes and a live birth) when using separate fits, or 28% (25 to 30) using the joint model: the proportion in the training set was 22%. Including the random effects, we predicted 23% (21 to 25) would do so.

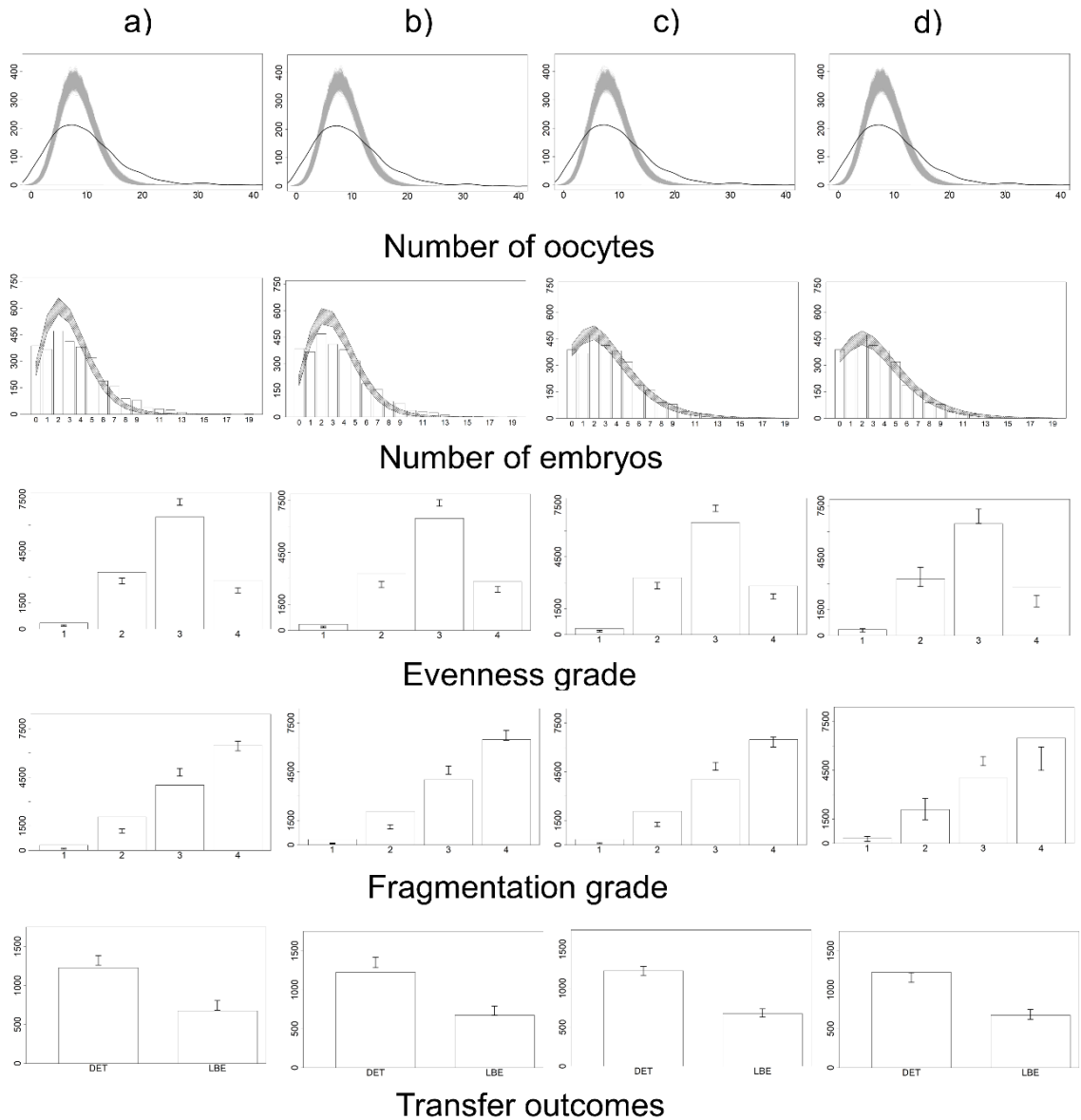

Figure 2: 95% posterior predictive intervals (shaded area or error bars) from a) pre treatment separate, b) pre treatment joint, c) dynamic separate, d) dynamic joint models, plotted with observed stage-specific responses. DET = double embryo transfer, LBE = live birth event.

## 3      Discussion

We have described modelling approaches for mixed, multilevel, sequential outcome data arising from a multistage treatment, and, using an application to routinely collected data, explored the feasibility of using these to make predictions in this context. All of the approaches described offer several advantages over those previously described, including the ability to incorporate mixed outcome types, without dichotomising into success or failure events, as well as responses defined at different levels of a multilevel data structure. The approaches are flexible enough to accommodate different combinations of response types and different covariates in the various submodels, according to the particular research question under consideration. The models can be fitted in freely available Bayesian software (38) without the need to write custom sampling algorithms.

We considered scenarios where predictions about cycle outcomes were to be made prior to treatment commencement, and also where predictions were to be made dynamically, using the outcomes of earlier treatment stages. The pre treatment scenario corresponds to patients deciding whether to undergo IVF. While prognostic models (40), as well as other forms of patient information relating to fertility treatment (41), have traditionally focussed exclusively on the chance of live birth for counselling prospective patients here, the current application highlights how a multivariate approach could be used to incorporate other important aspects of treatment, such as safety. Different patients may put different emphasis on potential costs and benefits of treatment, and the ability to predict a suite of outcomes with reasonable accuracy might enable improved personal decision making based on multiple factors. The dynamic approach uses information accrued up to that point to predict what will happen next. The application to real data suggested that knowledge of upstream outcomes could potentially improve prediction of subsequent stages, as has recently been described elsewhere (42), although an important caveat is that we included only a small number of patient characteristics as covariates. Nonetheless, we know that procedural outcomes such as numbers of eggs and embryos are predictive of live birth, and this fact has been leveraged in a model that uses information from a patient's previous IVF attempt (including stage-specific outcomes) to predict the live birth outcome in a subsequent attempt (43) . This differs from the situation we describe, where information accrues during a particular treatment attempt. The dynamic approach could be used to update patients on their prognosis as the cycle progresses. Patients and clinicians are known to do this in a subjective manner at present, and strong dynamic prediction might assist with expectation setting. There may be scope for clinicians to use the dynamic predictions to inform decisions relating to subsequent treatment stages (for example, relating to the number of embryos to transfer).

We would stress however that the application presented here is illustrative and exploratory, and not something we are endorsing as a clinical tool. We included only a handful of covariates in the models, and modelled continuous covariates as being linearly related. Many other variables are known to be predictive of the various stage-specific outcomes (44), and it would be necessary (although perhaps not sufficient) to incorporate these in order to attain good performance. Additionally, we have included six outcome variables in the current formulation, but these do not represent an exhaustive set of measurable outcomes in IVF. It may be useful to include supplementary (or alternative) outcomes in models intended for eventual clinical deployment. For example, in the present illustration, we do not distinguish between treatment failure due to transferred embryos not implanting in the uterine wall and failure due to implanted embryos not being sustained to term (ie: miscarriage). This may be an important distinction for many patients. We also do not distinguish singleton from twin births. The difference is clinically important, since twin pregnancies represent increased risk to the mother and infants. Close collaboration with patient and clinician stakeholders would be necessary to design a practical, useful, and relevant modelling framework, capturing the important outcome variables. Further complexity is introduced by variation in practice between IVF clinics. For example, different practices exist in relation to duration of embryo culture and the grading schemes used for embryo quality, with a recent interest in algorithmic approaches using parameters derived from video images captured within the incubator. The feasibility of including a variety of response types is indicated by this application.

Several important questions still require attention in the sequential outcomes context. These include questions about how to handle missing data, and whether there are special considerations relating to variable selection. By preselecting a small number of covariates that were universally recorded in the database, we have skirted these issues in the current example. The topic of evaluating performance of multivariate prediction models appears to be in its infancy. In the case of multiple binary outcomes, proposals have recently been made to extend the concepts of sensitivity, specificity, and predictive values to the multivariate setting (45). It remains to consider whether useful alternatives can be developed for the mixed outcome case. We have described graphical checks of predictive performance, and have described considerations relating to the denominator used for calculations in the sequential setting.

In both the pre treatment and dynamic examples, we did not observe clear advantages of joint modelling approaches compared to approaches based on modelling each outcome variable separately. We note that simulation studies are infeasible here, due to the computational expense of fitting the joint models. We have used an approach where submodels are connected using multivariate Normal latent variables. We have not considered whether alternative methods of joining the submodels, for example, using shared parameters, would be advantageous or feasible. Alternative proposals have been made using more flexible latent variable distributions, such as mixtures of Normals (46) or copula-based methods (47). It is possible that better results might be observed with (much) larger sample sizes.

While all of the models presented here can accommodate embryo-level response variables, relationships between these and other outcomes are estimated using the mean value (17). An undesirable consequence of this is the implicit assumption that the relationship between the evenness and fragmentation of an embryo is the same as the relationship between the evenness of an embryo and the fragmentation of another from the same patient (14). This could be relaxed by using latent representations of the embryo grading submodels and allowing the embryo-level residual terms to be correlated (12), (48). A related concern is the fact the models do not allow embryo-level responses to be included dynamically as covariates in the DET and LBE submodels without constructing a summary measure over a patient's embryos. The estimation of the effects of embryo characteristics on birth outcomes is complicated by the fact that if two embryos are transferred and only one implants, it is not known which of the two was successful. This partial observability problem motivates the use of embryo-uterine models which have been described from both Bayesian (49) and Likelihood (50) viewpoints. It remains to incorporate the embryo-uterine approach in the joint modelling approaches described here. We also note that the mean value might not be the best summary measure to use for the purpose of including the embryo gradings as covariates in the DET and LBE submodels, since the best one or two are selected for transfer. An alternative measure capturing the highest available grades might be more appropriate future applications of the methods. Alternatively, we could include the quality of the *transferred* embryos as additional embryo-level response variables in the model.

There are compelling arguments for considering multivariate outcome prediction in IVF, including the potential to realise multifactorial decision making incorporating potential risks of treatment, and, in the dynamic case, the scope to update prognosis for the purpose of expectation setting and to inform treatment decisions relating to the remaining treatment stages. We have introduced the idea using an application, and did not observe advantages of using a more complex (and computationally expensive) joint modelling approach. However, we stress the limits of what can be learned from a single example, and hope that this work may serve as a starting point for further discussion and development in this area.

**Declarations**

*Ethics approval and consent to participate*

This study used anonymised data collected for non-research purposes, and did not require ethical approval.

*Availability of data and materials*

The code used to perform these analyses is available at the Open Science Framework https://osf.io/pmrn3/ . The IVF clinic has not agreed to share the patient-level data analysed here.

*Competing interests*

JW declares that publishing research benefits his academic career.

*Funding*

JW is supported by a Wellcome Institutional Strategic Support Fund award [204796/Z/16/Z].

*Author contribution*

All authors conceived the idea, JW developed the methods and analysed the data, all authors contributed intellectual content, wrote the manuscript, and approved the final version.

**Figure legends**

Figure 1: Schematic of the IVF cycle

Figure 2: 95% posterior predictive intervals (shaded area or error bars) from a) pre treatment separate, b) pre treatment joint, c) dynamic separate, d) dynamic joint models, plotted with observed stage-specific responses. DET = double embryo transfer, LBE = live birth event.

| Variable | Summary |
|---|---|
| No of cycles started | 2962 |
| No of cycles where eggs mixed with sperm | 2861 |
| No of gradable embryos | 12911 |
| Number of embryo transfer procedures | 2501 |
| Age (years) | 33 |
| | 30 to 36 |
| | 21 to 43 |
| Partner Age (years) | 35 |
| | 32 to 39 |
| | 19 to 72 |
| Attempt Number | |
| 1 | 2132 (72%) |
| 2 | 659 (22%) |
| 3 | 147 (5%) |
| 4 | 4 (0%) |
| 5 | 20 (0%) |
| Number of eggs obtained per cycle started | 9 |
| | 5 to 13 |
| | 0 to 50 |
| Number of gradable embryos per cycle started | 3 |
| | 1 to 5 |
| | 0 to 19 |
| Number of embryos transferred per transfer procedure | |
| 1 | 1049 (42%) |
| 2 | 1452 (58%) |
| Live birth event per transfer procedure | |
| No | 1692 (68%) |
| Yes | 809 (32%) |

Table 1: Characteristics of the clinical dataset analysed in section 3. Median, interquartile range and range for continuous variables.

|  | Pre treatment | | Dynamic | |
| --- | --- | --- | --- | --- |
| **Number of oocytes submodel** | Separate | Joint | Separate | Joint |
| Intercept | 2.09 (2.07 to 2.12) | 2.10 (2.07 to 2.13) | 2.09 (2.07 to 2.12) | 2.09 (2.07 to 2.12) |
| Age (years) | -0.04 (-0.05 to -0.04) | -0.04 (-0.05 to -0.04) | -0.04 (-0.05 to -0.04) | -0.04 (-0.05 to -0.04) |
| Partner Age (years) | 0.01 (0.00 to 0.01) | 0.01 (0.00 to 0.01) | 0.01 (0.00 to 0.01) | 0.00 (0.00 to 0.01) |
| Attempt number: 1st | 0 | 0 | 0 | 0 |
| 2nd | 0.06 (0.01 to 0.12) | 0.06 (0.01 to 0.11) | 0.06 (0.01 to 0.12) | 0.08 (0.03 to 0.13) |
| 3rd | 0.17 (0.07 to 0.27) | 0.14 (0.04 to 0.23) | 0.17 (0.06 to 0.28) | 0.15 (0.06 to 0.24) |
| 4th or 5th | 0.02 (-0.24 to 0.29) | 0.04 (-0.20 to 0.27) | 0.02 (-0.24 to 0.26) | 0.16 (-0.07 to 0.38) |
| **Fertilization rate submodel** | | | | |
| Intercept | -1.04 (-1.06 to -1.01) | -0.95 (-0.98 to -0.92) | -1.04 (-1.07 to -1.01) | -0.96 (-0.99 to -0.93) |
| Age (years) | 0.01 (0.01 to 0.02) | 0.02 (0.01 to 0.03) | 0.01 (0.01 to 0.02) | 0.02 (0.01 to 0.02) |
| Partner Age (years) | 0.00 (-0.01 to 0.00) | -0.00 (-0.01 to 0.00) | 0.00 (-0.01 to 0.00) | 0.00 (-0.01 to 0.00) |
| **Embryo evenness submodel** | | | | |
| Intercepts (log odds of <=k): k=1 | -4.23 (-4.37 to -4.10) | -4.35 (-4.50 to -4.20) | -4.32 (-4.47 to -4.18) | -4.33 (-4.48 to -4.18) |
| K=2 | -1.28 (-1.37 to -1.19) | -1.38 (-1.48 to -1.29) | -1.37 (-1.47 to -1.27) | -1.37 (-1.47 to -1.26) |
| K=3 | 1.43 (1.34 to 1.51) | 1.33 (1.24 to 1.42) | 1.34 (1.25 to 1.44) | 1.35 (1.25 to 1.46) |
| Age (years) | 0.01 (-0.01 to 0.02) | 0.00 (-0.01 to 0.02) | 0.00 (-0.01 to 0.02) | 0.00 (-0.03 to 0.03) |
| Partner Age (years) | 0.01 (0.00 to 0.02) | 0.01 (0.00 to 0.02) | 0.01 (0.00 to 0.02) | 0.01 (0.00 to 0.02) |
| Sperm injected into egg | -0.21 (-0.32 to -0.10) | -0.27 (-0.38 to -0.16) | -0.28 (-0.39 to -0.16) | -0.26 (-0.37 to -0.14) |
| Number of oocytes | - | - | 0.09 (0.01 to 0.17) | 0.16 (-0.17 to 0.53) |
| Fertilisation rate | - | - | -0.16 (-0.23 to -0.08) | -0.47 (-0.71 to -0.21) |
| **Embryo fragmentation submodel** | | | | |
| Intercepts (log odds of <=k): k=1 | -4.91 (-5.10 to -4.74) | -5.11 (-5.29 to -4.92) | -5.06 (-5.24 to -4.88) | -5.06 (-5.25 to -4.88) |
| K=2 | -2.27 (-2.40 to -2.13) | -2.44 (-2.58 to -2.31) | -2.41 (-2.54 to -2.27) | -2.40 (-2.55 to -2.25) |
| K=3 | -0.15 (-0.27 to -0.03) | -0.33 (-0.46 to -0.20) | -0.29 (-0.42 to -0.17) | -0.28 (-0.42 to -0.14) |
| Age (years) | -0.02 (-0.05 to -0.00) | -0.03 (-0.05 to -0.01) | -0.03 (-0.05 to 0.00) | -0.06 (-0.1 to -0.02) |
| Partner Age (years) | 0.01 (-0.01 to 0.02) | 0.01 (0.00 to 0.02) | 0.01 (0.00 to 0.03) | 0.01 (-0.00 to 0.03) |
| Sperm injected into egg | -0.19 (-0.35 to -0.03) | -0.33 (-0.48 to -0.18) | -0.32 (-0.48 to -0.17) | -0.31 (-0.48 to -0.15) |
| Number of oocytes | - | - | 0.22 (0.10 to 0.33) | -0.18 (-0.67 to 0.29) |
| Fertilisation rate | - | - | -0.30 (-0.41 to -0.20) | -0.60 (-0.93 to -0.13) |
| **Double embryo transfer submodel** | | | | |
| Intercept | 0.11 (0.06 to 0.17) | 0.13 (0.07 to 0.19) | 0.13 (0.06 to 0.19) | 0.08 (0.02 to 0.14) |
| Age (years) | 0.02 (0.01 to 0.04) | 0.02 (0.01 to 0.04) | 0.02 (0.00 to 0.03) | -0.01 (-0.03 to 0.01) |
| Partner Age (years) | -0.01 (-0.02 to 0.00) | -0.01 (-0.02 to 0.00) | 0.00 (-0.02 to 0.00) | -0.00 (-0.01 to 0.01) |
| Attempt No: 1st | 0 | 0 | 0 | 0 |
| 2nd | 0.27 (0.15 to 0.39) | 0.23 (0.11 to 0.35) | 0.25 (0.13 to 0.37) | 0.27 (0.15 to 0.38) |
| 3rd | 0.46 (0.23 to 0.70) | 0.44 (0.19 to 0.67) | 0.47 (0.23 to 0.72) | 0.53 (0.31 to 0.75) |
| 4th or 5th | 0.72 (0.15 to 1.31) | 0.60 (0.06 to 1.17) | 0.63 (0.10 to 1.21) | 0.69 (0.16 to 1.25) |
| Number of oocytes | - | - | -0.06 (-0.13 to -0.03) | -0.25 (-0.50 to 0.01) |
| Fertilization rate | - | - | -0.02 (-0.09 to 0.05) | -0.34 (-0.51 to -0.10) |
| Embryo evenness | - | - | -0.14 (-0.20 to -0.08) | -0.09 (-0.15 to -0.02) |
| Embryo fragmentation | - | - | -0.12 (-0.18 to -0.06) | -0.05 (-0.12 to 0.03) |
| **Live birth event submodel** | | | | |
| Intercept | -0.46 (-0.51 to -0.41) | -0.49 (-0.54 to -0.44) | -0.56 (-0.64 to -0.47) | -0.39 (-0.73 to -0.04) |
| Age (years) | -0.02 (-0.04 to 0.00) | -0.02 (-0.04 to -0.01) | -0.01 (-0.03 to 0.00) | -0.03 (-0.05 to -0.01) |
| Partner Age (years) | -0.01 (-0.02 to 0.00) | -0.01 (-0.02 to 0.00) | -0.01 (-0.02 to 0.00) | 0.00 (-0.01 to 0.00) |
| Number of oocytes | - | - | 0.04 (-0.03 to 0.12) | -0.13 (-0.31 to 0.08) |
| Fertilization rate | - | - | 0.14 (0.07 to 0.21) | -0.16 (-0.39 to 0.12) |
| Embryo evenness | - | - | 0.07 (0.01 to 0.13) | 0.03 (-0.04 to 0.10) |
| Embryo fragmentation | - | - | 0.04 (-0.02 to 0.10) | 0.04 (-0.04 to 0.11) |
| DET | - | - | 0.11 (0.01 to 0.22) | -0.13 (-0.66 to 0.43) |

Table 2: Estimated regression coefficients (95% CIs) from the fitted models